\documentclass[useAMS,usenatbib]{mn2e}
\usepackage{lscape}
\usepackage{rotating}
\usepackage{amssymb}
\usepackage{float}
\usepackage{txfonts}
\usepackage{natbib}

\def\cm3{\hbox{cm$^{-3}$}}
\def\iras{\hbox{IRAS~13224-3809}}
\def\1h0707{\hbox{1H~0707-495}}
\def\ngc4051{\hbox{NGC~4051}}
\def\mcg63015{\hbox{MCG~6-30-15}}

\input psfig.sty
\input epsf.sty
\usepackage{graphicx}
\usepackage{epsfig}
\title[Timing/spectral study of \iras]
{A combined timing/spectral study of \iras\ using {\it XMM-Newton} data}

\author[Caballero-Garc\'{i}a et~al.]{M.~D. Caballero-Garc\'{i}a$^{1}$\thanks{E-mail:
garcia@asu.cas.cz}, I.~E. Papadakis$^{2,3}$, M. Dov\v{c}iak$^{1}$, M. Bursa$^{1}$, \newauthor
J. Svoboda$^{1}$, V. Karas$^{1}$
\\
\\
$^{1}$ Astronomical Institute of the Czech Academy of Sciences, Bo\v{c}n\'{\i}~II~1401, CZ-14100~Prague, Czech~Republic \\
$^{2}$ Department of Physics and Institute of Theoretical and Computational Physics, University of Crete, 71003 Heraklion, Greece \\
$^{3}$ IA, FORTH, Voutes, GR-7110 Heraklion, Greece \\
}

\date{Accepted. Received; in original form}
\pagerange{\pageref{firstpage}--\pageref{lastpage}}

\begin{document}
\maketitle
\label{firstpage}
\begin{abstract}
We present the results from an X-ray variability study of IRAS 13224-3809. This is probably the best source for X-ray reverberation studies since it is X-ray bright, extremely variable, and it has been extensively observed with
{\it XMM-Newton}. We used all the archival {\it XMM-Newton} data from the three EPIC cameras (to increase the signal-to-noise) and, given the many observations of the source, we were able to compute the time-lags spectra in three different flux levels/periods. We fitted the time-lags and energy spectra, simultaneously, using a new X-ray reverberation code which computes the time dependent reflection spectra of the disc as a response to an X-ray flash
from a point source located on the axis of the black-hole (BH) accretion disc (lamp-post
geometry). To the best of our knowledge, this is the first time for an AGN that both time-lags and energy spectra are fitted by a model simultaneously in different flux periods. The model fits in the case when the BH is rapidly rotating are significantly better than the model fits in the case of a Schwarzschild BH. This result strongly favours the hypothesis of a rotating central BH in this source. We also detect significant variations in the height of the X-ray corona. The X-ray height appears to increase from $\sim 3 - 5$ gravitational radii when the X-ray luminosity is of the order of $\sim 1.5-3$ percent of the Eddington limit, up to $\sim 10$ gravitational radii, when the luminosity doubles. 
\end{abstract}

\section{Introduction}
\label{sec:intro}

The most commonly accepted scenario for Active Galactic Nuclei (AGN) postulates that they are powered by accretion of matter onto a central super massive 
black hole (BH). AGN are strong X-ray emitters, probably via inverse Compton scattering of the optical/UV photons which are emitted from the accretion disc 
by hot electrons. The resulting X-ray energy spectrum has a power-law shape up to an energy characteristic of the electron temperature. 

The X-ray source is believed to be located in the innermost part of the accretion disc since the X-ray luminosity is a sizeable fraction of the bolometric 
luminosity. It is also small in size, with radii not much larger than a few gravitational radii (${\rm r}_{\rm g}={\rm GM_{\rm BH}}/{\rm c}^2$; $\rm M_{BH}$ is the 
mass of the central BH). This is based on the high amplitude, fast variations that are observed in the X-ray emission from these objects. This is also supported by 
recent monitoring observations of several lensed quasars performed in the optical, UV, and X-ray bands (e.g. \citealt{chartas16}). 

In bright AGN, with strong ``big blue bump", the disc probably extends down to the inner stable circular orbit, which implies that the X-ray source will be located 
at some height on top of it. If the X-ray emission is isotropic then it is unavoidable that the X-rays will illuminate the accretion disc. Part of the X-rays will be absorbed 
and will increase the temperature of the disc, and part of them will be ``reflected", giving rise to the so called X-ray reflection spectrum. The main features of this 
spectrum are: an excess of emission at energies below $\sim 1-1.5$ keV (i.e. in the ``soft" energy band), the presence of an iron line at $\sim 6.4-7$ keV, and the 
broad ``Compton hump" at energies $\sim 20-50$ keV. Depending on the height of the X-ray source, the inner disc radius and its ionisation, these features will be 
affected by relativistic effects (e.g. the iron line will be broadened and the soft band emission will appear as a broad, featureless emission component). These features 
have been observed in the X-ray spectra of many AGN, although they may be explained in more than one way. 

If X-ray irradiation of the inner disc takes place in AGN, we expect to observe correlated variations in the X-ray continuum and the reflection components. The latter 
will be delayed, depending on the distance between the X-ray source and the accretion disc. Time delays between the soft band and the X-ray continuum variations were 
first reported in Ark 564 \citep{mchardy07}, and were conclusively detected in 1H~0707-495 \citep{fabian09}. Since then, such delays have been detected in many more 
objects \citep{emmanoulopoulos11,demarco13,kara16}.

Theoretical modelling of the observed time-lags vs. frequency (i.e. the ``time-lags spectra") and of the 
time-lags, at a fixed frequency, as a function of energy have been performed a few times in the past \citep{cackett14,emmanoulopoulos14,epitropakis16a,chainakun15,chainakun16,chainakun17}. Recently 
\citet{caballerogarcia18} studied the time-lags spectra of three AGN. They used a model which assumes a point-like X-ray source located at a height $h$ above 
the BH (i.e. the so call ``lamp-post" geometry; Matt et al. 1991) to fit the observed time-lags between the 0.3-1 and 1-10\,keV bands. The model fitted the data well, and the 
best-fit height was $\approx4\,{\rm r}_{\rm g}$, on average, irrespective of the BH spin considered and the disc ionization. They also detected wavy-like residuals in the 
time-lags spectra at high frequencies, which suggest that the lamp-post model, in its simplest version, should be modified. Either the X-ray source is extended, or the height 
of the source varies with time. 

IRAS~13224-3809 is a radio-quiet, X-ray bright, Narrow Line Seyfert 1 (NLSy1) galaxy, located at
a luminosity distance of 310 Mpc (NED \footnote{The NASA/IPAC Extragalactic Database (NED) is funded by the National Aeronautics and Space Administration and operated by 
the California Institute of Technology.} database). It has been extensively studied in X-rays as it is one of the most variable Seyfert galaxies in this band, both in flux 
and spectral shape (i.e. \citealt{boller97,boller03,dewangan02,gallo04,ponti10,fabian13,chiang15,parker18,pinto18,jiang18}). {\it XMM-Newton} observed \iras\ in 2002, 2011 and 
2016 for $\sim 1$\,day, 500\,ks and 1.5\,Ms, respectively. When combined together, these are among the longest {\it XMM-Newton} observations of a single AGN. Given the 
continuous, fast and large-amplitude variability of the source, they constitute a valuable data set to study the variability properties of the source in detail. 

 \citet{kara13}, \citet{emmanoulopoulos14} and \citet{chainakun16} used the 2011 observation to study the time-lags spectrum versus frequency and energy of the source. \citet{kara13} 
studied the time-lags versus frequency spectra and found frequency and absolute amplitude variations between flaring and quiescent periods. They suggested that the X-ray source may 
be located closer to the accretion disc in the lower-flux than during the high flux period. \citet{emmanoulopoulos14} fitted the time-average time-lags versus frequency spectrum in the 
lamp-post geometry, and found that the X-ray source is located at $\sim 3\,{\rm r}_{\rm g}$ above the central BH. \citet{chainakun16} fitted both the time average time-lags versus 
energy and the energy spectrum with a similar model. They also found a small X-ray source height (of the order of $2.0\,{\rm r}_{\rm g}$). 
 
Recently, \citet{alston19} used the full data set to study the variability properties of the source. They found strong evidence for non-stationarity. They showed that the rms-flux 
relation is not linear, and the power-spectrum density (PSD) normalisation increases with decreasing source flux, while the low-frequency peak moves to higher frequencies. \citet{alston20} used 
(almost) the same data to fit the time-lags spectra in time scales of the order of a day. They used the same model that G18 used, and found variations that could be explained if the height of 
the X-ray corona increases with increasing luminosity. 
 
We present the results from a simultaneous spectral/timing study of \iras. Like \citet{alston20} we use the full {\it XMM-Newton} data set but we consider both EPIC-pn and MOS data 
to reach the maximum possible signal-to-noise (S/N) ratio for the timing analysis. We also use an improved version of the model that G18 and \citet{alston20} used and 
we fit, simultaneously, {\it both} the time-lags versus frequency spectra and the energy spectrum. Our approach is similar to \citet{chainakun15,chainakun17} employed for another 
AGNs. Compared to the later authors, we use a longer data set, and we investigate the spectral/timing properties of the source in different flux levels (hereafter referred to 
as ``periods" of time of \iras\ with similar fluxes). To the best of our knowledge, this is the first time that both the X-ray time-lags and energy spectrum of an AGN, in different flux 
periods, are fitted with a model that takes into account all the relativistic effects and the ionization radial profile of the disc. 

\section{Data reduction}  \label{sec:analysis}

We considered the three long {\it XMM-Newton} observations of {\iras} in 2002, 2011 and 2016 (Tab.~\ref{table1} lists the observations log). The EPIC pn and MOS 
cameras were operating in the {\it Full Frame} and the {\it Large Window} imaging modes, respectively, during the 2002 and 2011 observations. They both operated in the {\it Large Window} mode 
during the 2016 observation. The {\it Thin} filter was used in the 2011 and 2016 observations, and the {\it Medium} filter in the 2002 observation. We processed the data from the 
EPIC-pn \citep{struder01} and the two MOS \citep{turner01} cameras using the {\tt Scientific Analysis System} (SAS) v. 16.1.0 \citep{gabriel04}.

EPIC-pn and MOS1, MOS2 source data were extracted using circular regions with a radius of 800 pixels ($40^{\prime\prime}$) centred on the source coordinates as listed on the NASA/IPAC 
Extragalactic Database (${\rm RA}=13\,h25\,m19.4\,s$, ${\rm Dec}=-38\,d24\,m52.6\,s$) which were the same for the timing and the spectral analysis for each observation. We kept only events 
with {\tt PATTERN}${\le}4$ and {\tt FLAG}$=0$ for the EPIC-pn data, and {\tt PATTERN}${\le}12$ and {\tt FLAG}$=0$ for the MOS data. We assessed whether the observations are affected by 
pile-up using the SAS task {\tt epatplot}, and we did not detect significant pile-up effects in any of them. Background data were extracted from large rectangular boxes far enough from 
the location of the source, but remaining within the boundaries of the same CCD chip (they were also the same for each observation for the timing and the spectral analysis, respectively). 

We used the {\it Science Analysis System} (SAS) tool {\tt evselect} to produce the source and background light curves in the 0.3-1\,keV and 1-4\,keV energy bands (the ``soft" and ``continuum" 
bands, respectively), using a time bin size of 100\,s for all three cameras. Background subtracted light curves were produced using the SAS tool {\tt epiclccorr}. We checked for background flaring 
events, and we removed data points where the background count rate was half or higher than the source rate, in both energy bands. 

\begin{table}
 \centering
 \caption{\textit{XMM-Newton} observations log. }
\label{table1}
\begin{tabular}{c c c  c     }     
\hline
Date Obs. & Obs. ID & Duration \\ 
 (dd/mm/yyyy)       & &  (s)   \\ 
\hline
19/01/2002   &  0110890101 & 64019  \\ 
19/07/2011   &  0673580101 & 133039 \\ 
21/07/2011   &  0673580201 & 132443 \\ 
25/07/2011   &  0673580301 & 129438 \\ 
29/07/2011   &  0673580401 & 134736 \\ 
08/07/2016   &  0780560101 & 141300 \\ 
10/07/2016   &  0780561301 & 141000 \\ 
12/07/2016   &  0780561401 & 138100 \\ 
20/07/2016   &  0780561501 & 140800 \\ 
22/07/2016   &  0780561601 & 140800 \\ 
24/07/2016   &  0780561701 & 140800 \\ 
26/07/2016   &  0792180101 & 141000 \\ 
30/07/2016   &  0792180201 & 140500 \\ 
01/08/2016   &  0792180301 & 140500 \\ 
03/08/2016   &  0792180401 & 140800 \\ 
07/08/2016   &  0792180501 & 138000 \\ 
09/08/2016   &  0792180601 & 136000 \\ 
\hline
\end{tabular}
\end{table}

\begin{figure*}
\centering
 \includegraphics[bb=59 22 570 735,width=0.69\textwidth,angle=270,clip]{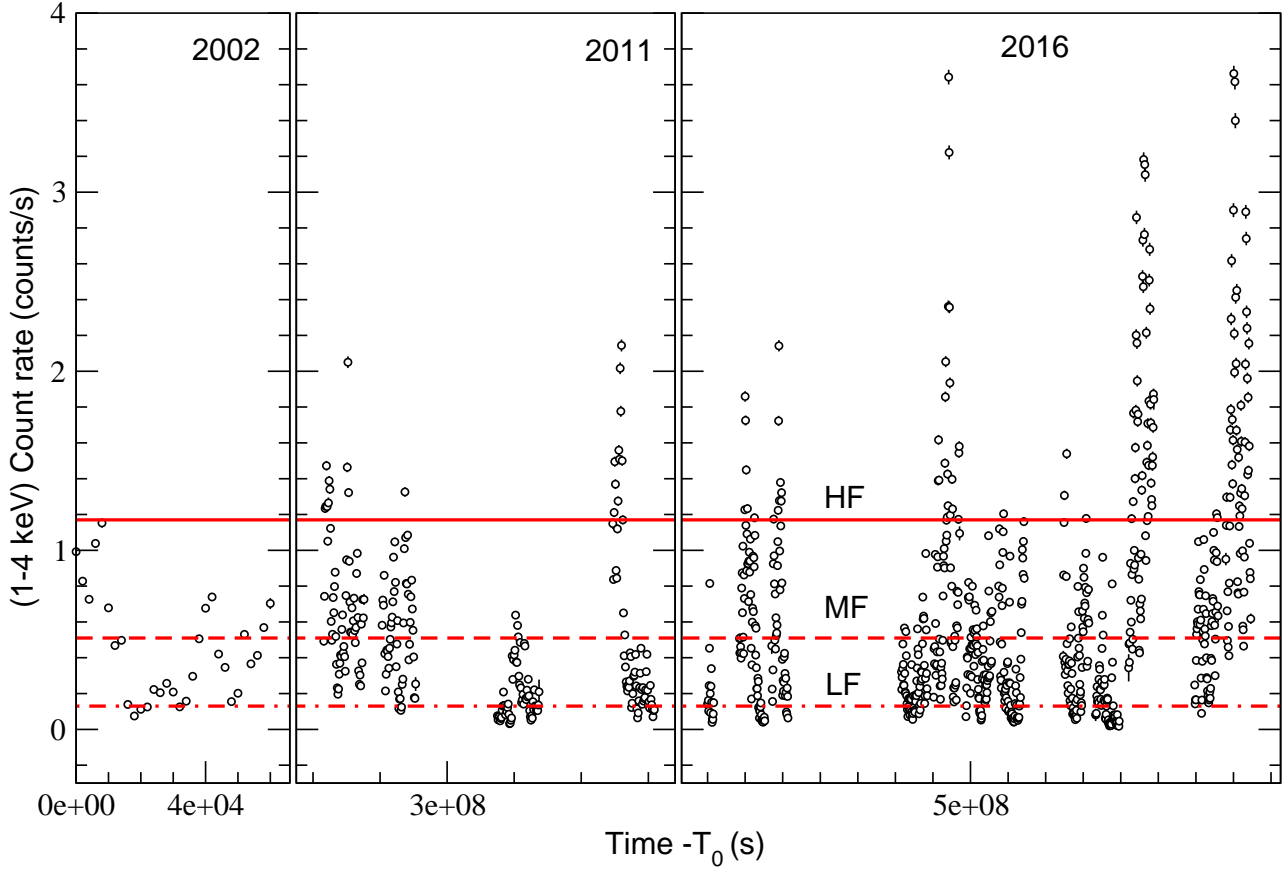}
\caption{ The EPIC pn+MOS, 1-4\,keV band light curve, binned in 2 ks. Time is measured in seconds since the start of the 2002 observation. The solid, dashed and dot-dashed horizontal lines indicate the count rate limits of the ``HF", ``MF" and ``LF" periods (see text for details). }
\label{fig:lc1-4}
\end{figure*}

We summed the MOS1$+$MOS2 light curves (hereafter referred to as the MOS light curves) using the {\tt lcmath} in FTOOL. Then we added the pn with the MOS light curve, producing 
the final light curves in the two energy bands. Before adding the light curves, we divided one over the other, to check whether their ratio was stable and consistent with 
a constant. This was the case almost always (except for a few, short, light curve parts at the beginning and/or the end of each observation, which we ignored). 

The final light curves cover periods when both instruments were 
operating. They contain a small number of missing points ($\lesssim5\%$ of the total number of points in each light curve). They are randomly distributed throughout the duration of an 
observation, or appear in groups of ${\lesssim}10$ points (maximum). We replaced the missing points by a linear interpolation, with the addition of the appropriate Poisson noise.

In Fig.~\ref{fig:lc1-4} we show the final, pn+MOS, 1-4\,keV light curve binned to 2\,ks. In the following section we describe the way we estimated the time-lags in the three ``flux periods", namely 
the ``High-flux (HF)", the ``Medium-flux (MF)", and the ``Low-flux (LF)" periods. The horizontal lines in Fig.~\ref{fig:lc1-4} at 0.13, 0.51 and 1.17\,${\rm s}^{-1}$ indicate the count rate limits that we used to define these periods. We used the 1-4 keV band to define the flux periods, because this band should be more representative of the primary X-ray emission. 
 
For the energy spectral analysis we used the EPIC-pn camera only, for simplicity and to avoid possible issues due to cross-calibration effects between the pn and MOS cameras. We chose 
the EPIC-pn camera because it has a larger effective area (i.e. double) than each one of the MOS cameras. We built response matrix functions with the SAS tasks {\tt rmfgen} and {\tt arfgen}. We 
used {\tt evselect} to produce three EPIC-pn energy spectra for each observation identifier (OBSID), i.e. one for each LF, MF and HF periods, using the start and end time of all the light 
curve segments that we used to construct the LF, MF and HF light curves. We combined the individual spectra using the {\tt epicspeccombine} task, and we created the final LF, MF and HF, source 
and background spectra. The spectra were rebined using the {\tt grppha} task to have at least 200 counts for each background-subtracted spectral bin.

\section{time-lags estimation}  \label{sec:lags}

We used the method proposed by \citet[][EP16 hereafter]{epitropakis16b} to estimate the time-lags between the 0.3-1\,keV and 1-4\,keV light curves. A detailed description of the method is 
given in EP16. We summarise below the main points, for consistency and clarity. 

First, we divided the light curves into segments of duration $T=10$\,ksec. Each segment was placed into the HF, MF or LF periods, if its mean count rate in the 1-4\,keV band was larger than 1.17, between 0.51-1.17 and between 0.13-0.51 counts/s, respectively (see Fig.~\ref{fig:lc1-4}). There are $m=80, 60$ and 27 segments in the LF, MF and HF periods, respectively. We calculated the cross-periodogram of all the HF, MF and LF segments at frequencies $\nu_p=p/(N\Delta t)$ ($p=1,...,N/2, N=100$, and $\Delta t=100$\,s, in our case). Our cross-spectrum 
and time-lag spectrum estimates are, 
\noindent
\begin{equation} \label{eq:eq1}
\hat{C}_{xy}(\nu_p)=\frac{1}{m}\sum_{k=1}^{m}I^{(k)}_{xy}(\nu_p),
\end{equation}
\noindent
and
\noindent
\begin{equation} \label{eq:eq2}
\hat{\tau}_{xy}(\nu_p)\equiv\frac{1}{2\pi\nu_p}\mathrm{arg}[\hat{C}_{xy}(\nu_p)],
\end{equation}
\noindent
respectively ($I^{(k)}_{xy}(\nu_p)$ is the cross-periodogram of the $k-$th segment at frequency $\nu_p$). The error of of $\hat{\tau}_{xy}(\nu_p)$ is given by: 
\noindent
\begin{equation} \label{eq:eq3}
\sigma_{\hat{\tau}}(\nu_p)\equiv\frac{1}{2\pi\nu_p}\frac{1}{\sqrt{2m}}\sqrt{\frac{1-\hat{\gamma}^2_{xy}(\nu_p)}{\hat{\gamma}^2_{xy}(\nu_p)}},
\end{equation}
\noindent
where,
\noindent
\begin{equation} \label{eq:eq4}
\hat{\gamma}^2_{xy}(\nu_p)\equiv\frac{|\hat{C}_{xy}(\nu_p)|^2}{\hat{P}_x(\nu_p)\hat{P}_y(\nu_p)}.
\end{equation}
\noindent
$\hat{P}_x(\nu_p)$ and $\hat{P}_y(\nu_p)$ are the traditional periodograms of the two light curves, which are also calculated by binning over $m$ segments, and $\hat{\gamma}^2_{xy}(\nu_p)$ 
is the sample coherence function (defined on the interval $[0,1]$). 

\begin{figure}
\centering
\includegraphics[bb=7 17 560 760,width=9cm,angle=0,clip]{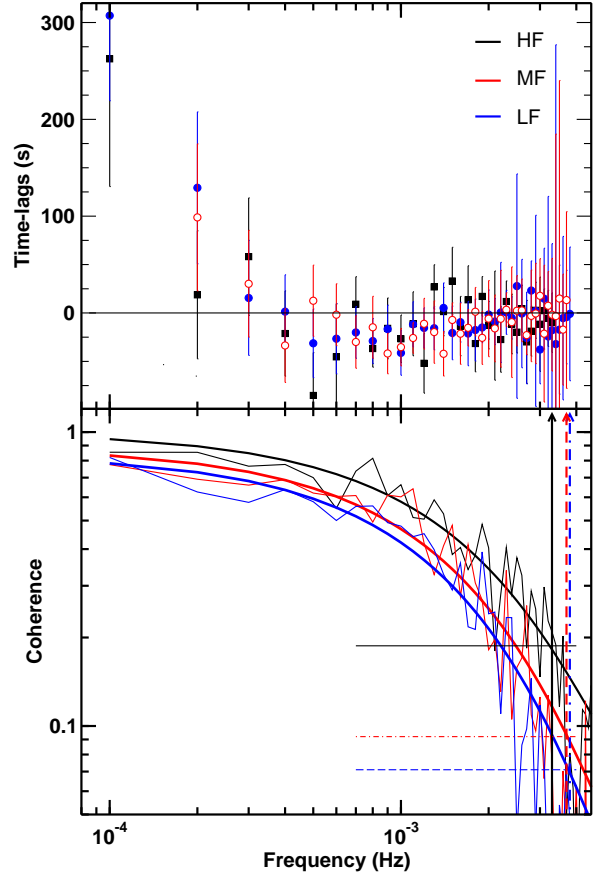}
\caption{The HF, MF, and LF time-lags spectra (top panel; filled squares, open and filled circles, respectively) and the respective coherence functions (bottom panel; data are connected by black, red and blue solid lines, from the top to the bottom, respectively). The solid black, dashed read and dot-dashed blue vertical lines in the bottom panel indicate the highest frequency up to which we can estimate the HF, MF and LF time-lags, respectively (see text for details). }
\label{fig:timelags}
\end{figure}

Fig.\,\ref{fig:timelags} shows the HF, MF and LF time-lags spectra (top panel) and the respective coherence functions (bottom panel). 
The time-lags in the three flux periods appear rather similar. They become negative at around $\sim 3\times 10^{-4}$\,Hz and then broadly approach and cross zero at frequencies higher than 0.002 Hz. The coherence decreases to zero at high frequencies due to the Poisson noise effects (but also due to intrinsic reason; see EP17). The loss of coherence at high frequencies can be reasonably well approximated by an exponential function of the form (EP16),
\noindent
\begin{equation} \label{eq:eq5}
\hat\gamma^2_{xy}(\nu)=\left(1-\frac{1}{m}\right)\mathrm{exp}[-(\nu/\nu_0)^q]+\frac{1}{m}.
\end{equation}
\noindent

The solid, smooth black, blue and red curves in the bottom panel of Fig.\,\ref{fig:timelags} show the best-fit model lines to the HF, MF and LF coherence. The vertical lines 
in the same panel indicate the frequency at which the best-fit coherence functions are equal to $1.2/(1+0.2m)$. This is the maximum frequency, $\nu_{\mathrm{max}}$ , at which we can reliably 
estimate time-lags. Above $\nu_{\mathrm{max}}$, the time-lag estimates are biased (they converge to zero, irrespective of their true value) and eq.\,\ref{eq:eq3} underestimates their 
true error. In our case, $\nu_{max,{\rm HF}}=3.3\times 10^{-3}$ Hz, and $\nu_{max,{\rm MF, LF}}\sim 3.7\times 10^{-3}$\,Hz. So we decided to keep the time-lags spectra up to $3.3\times 10^{-3}$ Hz only, for all flux periods. 

\begin{figure}
\centering
\includegraphics[bb=35 350 540 770,width=9cm,angle=0,clip]{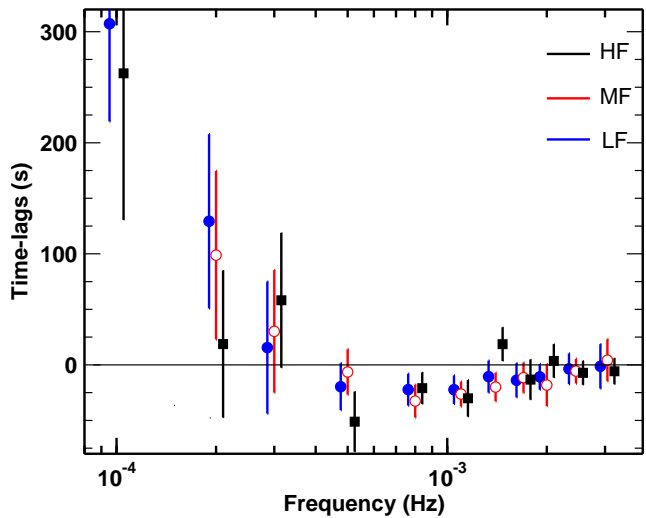}
\caption{The binned HF, MF, and LF time-lags spectra (filled squares, open and filled circles, respectively). The HF and the LF time-lags are shifted in frequencies by a factor of 1.1 and 0.9, respectively, for clarity reasons.  }
\label{fig:timelagssmooth}
\end{figure}

In order to increase the signal-to-noise ratio of the time-lags, we smoothed them by binning consecutive time-lag estimates. We did not bin the three lowest frequency points (to constrain the low frequency, positive, hard lags as much as possible) and we used bins of size 3 and 6 at higher frequencies. The resulting time-lags spectra appear in Fig.\,\ref{fig:timelagssmooth}. Soft, negative time-lags are clearly observed for almost a decade in frequency (from $\sim 3\times 10^{-4}$ up to 0.002\,Hz). We used the binned time-lags shown in this figure in the model fitting process we present below.

\section{Model fitting and results}  \label{sec:fitting}

\subsection{Initial considerations} 
\label{model}

Due to its long {\it XMM-Newton} observations, \iras\ is the only AGN that we can compute the time-lags at various flux periods, with a reasonable accuracy. This possibility offers us 
additional information that we can use to determine important physical parameters of the source, like the BH spin and the source height for example. However, even in this case, this is 
not an easy task. This is due to the fact that the observed time-lags depend on many parameters, like the BH mass, source height (and size, in practice), disc ionization, iron abundance 
and inclination of the disc, BH spin, etc. In principle, we could also estimate the 5-7 vs. 1-4\,keV band time-lags (i. e. the time-lags between the iron-line and the continuum 
variations) and fit, simultaneously, both the soft and the iron line time-lags spectra, to determine the model parameters with higher accuracy. However, this cannot be done in \iras. The 
source has a steep X-ray spectrum, and the count rate at energies above $\gtrsim 5$\,keV is very low. As a result, it is not possible to compute the iron-line time-lags spectra with any 
reasonable accuracy, even if we consider the full data set (if we split the light curves in various flux periods). 

There is a consensus in the scientific X-ray community that (whenever it is possible) it is important to study spectral and timing data together to consistently and accurately determine the
physical parameters of the disc/corona geometry. We decided to follow this approach\footnote{This decision was made after performing timing and spectroscopy separately and confirming that the results 
are consistent with the ones presented in this paper.} and present it here. So we fit the time-lags versus frequency simultaneously with the energy spectrum of the source. 

The full band X-ray spectrum of \iras\ is notoriously complicated (see e.g. \citealt{jiang18} for an analysis of the X-ray spectrum in the full 0.3-10\,keV {\it XMM-Newton} pass-band 
during the 2016 observation). In addition to the continuum and the X-ray reflection components, spectral features due to warm absorbing material as well as extra broad spectral components 
in the soft band can also be observed. In order to keep our modelling as less affected from these additional components as possible, we decided to fit the time-lags spectra plotted 
in Fig.\,\ref{fig:timelagssmooth} together with the HF, MF and LF energy spectra in the 3-10\,keV  band. According to Fig. 10 of \citet{jiang18}, the main components in the energy spectrum of 
the source at energies $3-10$\,keV are the continuum and the reflection component(s) with the prominent iron line feature. So, the time-lags between the 0.3-1 vs. 1-4\,keV band light curves 
hold information regarding the reverberating X-ray reflection spectrum in the soft band while the energy spectrum in the 3-10\,keV band is representative of the reflection signal in the iron line band. 

\subsection{The model fitting process}

We fitted both the time-lags and the energy spectra, simultaneously, in the three flux periods. We ignored the data in the energy range $7.5-8.3,7.5-8.3$ and $7.5-8.1$\,keV for the LF, MF and HF 
energy spectra, respectively, due to the absorption features that appear in these bands (e.g. \citealt{parker17,chartas18}). 

We used the {\tt KYNXILREV}\footnote{https://projects.asu.cas.cz/stronggravity/kynreverb} model to fit the time-lags spectra. This is almost identical to the {\tt KYNREFREV} model that 
we used in G18, with the exception that, in this version, re-processing of the ionised disc is computed using the {\tt XILLVER} tables \citep{garcia10,garcia13}. As for fitting the 
energy spectra, we used the {\tt KYNXILLVER} model, which is part of the {\tt KYN}\footnote{https://projects.asu.cas.cz/stronggravity/kyn} set of models first presented in \citet{dovciak04} and
later in \citet{dovciak14b}. {\tt KYNXILREV} can fit both time-lags and energy spectra. However, {\tt KYNXILLVER} is much faster in fitting energy spectra, and since both models are 
identical (in the assumed geometry, relativistic and X-ray reflection modelling) we decided to use the later for fitting the energy spectra. 

Both {\tt KYNXILREV} and {\tt KYNXILLVER} assume the so called lamp-post geometry. The X-ray source is assumed to be point-like, 
located on the axis of the black-hole accretion disc system, at a height $h$ above the BH. The X-ray continuum spectrum is of the form: 
${\rm F}(\rm E){\propto}{\rm E}^{-{\Gamma}}\mathrm{e}^{-{\rm E}/{\rm E}_{\rm c}}$, where ${\rm E_c}= 300$\,keV. The X-ray source is assumed to emit isotropically, in its 
own frame, and the models take into account all the relativistic effects in the propagation of light from the primary source to the disc and to the observer, and from the 
disc to the observer. The models assume a thin, Keplerian disc, co-rotating with the BH. The inner disc edge is equal to the radius of the innermost stable circular 
orbit, ${\rm r}_{\rm ISCO}$, and the outer radius, ${\rm r}_{\rm out}$, is fixed at $10^{3}\,{\rm r}_{\rm g}$. The inner disc radius is determined by the BH spin. The ionisation 
state of the disc changes with radius and depends on the illuminating pattern (of the continuum as it 
illuminates the disc) as well as on the density radial profile of the disc. In order to model the reflection spectrum we used the {\tt XILLVERD} tables \citep{garcia16} to be able 
to allow the disc density to be a variable parameter during the model fits. However, we assumed (for simplicity) a constant density radial profile.

One of the main physical parameters of both models is the BH mass, M$_{\rm 8}$ (in units of $10^8{\rm M}_{\odot}$). We kept it fixed to 
$2{\times}10^{6}\,{\rm M}_{\odot}$ during the model fitting\footnote{This BH mass estimate is based on the results from the recent sudy of X-ray reverberation properties of \iras\
\citep{alston20} and the results of the study of its PSD from \citet{alston19} using the relationship between PSD break timescale, black hole mass and luminosity from \citet{mchardy06}.}. This
was made in order to be able to determine the height of the X-ray source as accurately as possible. The X-ray source height and the BH mass affect the time-lags in 
a similar way. Therefore, if the BH mass in \iras\ is larger/smaller than the adopted BH mass then the source height (in gravitational radii) should be smaller/larger than our 
best-fit estimates, proportionally. However, the BH mass value cannot affect any relative changes between the best-fit $h$ values in the three flux periods. 
 
Another important physical parameter is the BH spin, $a$, which defines ${\rm r}_{\rm ISCO}$. When measured in geometrized units, $a$ can attain any value between zero for a 
Schwarzschild BH (in which case ${\rm r}_{\rm ISCO}=6\,{\rm r}_{\rm g}$), and 1 in the case of a co-rotating disc around a rapidly spinning BH (in which 
case ${\rm r}_{\rm ISCO}=1\,{\rm r_g}$). Although the light curves from \iras\ are very long, they are still not good enough to constrain the spin of the source with a high 
accuracy (even if we fit simultaneously time-lags and energy spectra). For that reason, following G18, we decided to fit the data in the two extreme cases of $a=0$ or $a=0.99$. In 
this way, we wish to investigate whether the existing spectral/timing data can discriminate between a non-rotating and a rapidly rotating BH. We used a value for the inclination of the accretion
disc of ${\theta}_{0}=45$\,deg., i.e. an intermediate value.

The {\tt KYNXILREV} model includes a phenomenological, power-law-like prescription for the hard (i.e. positive) time-lags at low frequencies, of the form:
\begin{equation}
\tau(\nu)={\rm A}\nu^{-s}.
\end{equation}
The power-law-like assumption of the low-frequency positive lags is a good description of the observed time-lags in various energy bands of many AGN (e.g. \citealt{epitropakis17}). 

During the fitting process we obtain the observed primary isotropic fluxes in the 2-10\,keV energy range, in units of the Eddington 
luminosity (${\rm L}_{\rm EDD}=2.5{\times}10^{44}\,{\rm erg}\,{\rm s}^{-1}$ for the adopted BH mass shown in Tab.~\ref{log_results5}). Additionally, we fix the spectral 
normalization value to $1.04{\times}10^{-5}$. The latter is the 
normalization of the {\tt kynxillver} model (defined as ${\rm norm}=1/{\rm D}^2$, where D is the
distance to the source in Mpc). This value was set according to the luminosity distance of the source, i.e. 310\,Mpc. 

We note that we also tested for the effects of possible absorption by an extended outflow \citep{pinto18} to the validity of our fit results. If the absorption 
is by highly ionized material and it is uniform at energies above 3 keV, then it will mainly affect the estimation of the intrinsic X-ray luminosity. To verify 
this, we set the {\tt kynxillver} normalization to be equal to $5.2\times 10^{-6}$. This is half the value we mentioned in the previous paragraph, and it corresponds 
to the case when the observed flux is halved due to absorption. We fitted the data (time-lags and energy spectra), and the quality of the best-fit turned to be the 
same as before (i.e. the best-fit $\chi^2$ values for the two spin parameters were almost identical to the values listed in Tab.~\ref{log_results5}). The best fit 
values were also very similar to the values listed in Tab.~\ref{log_results5}, except of L$_{2-10}/$L$_{\rm Edd}$, which was doubled, and the disc density that 
was approximately 1.5 times larger. 

The model parameters that we let free to vary during the model fits are the following: the iron abundance (with respect to the solar, Z/Z${_\odot}$), the disc density, the height of the X-ray 
source ({\it h}, in units of ${\rm r}_{\rm g}$), the observed primary X-ray flux in the 2-10\,keV band (L$_{2-10}$, in units of the Eddington luminosity, ${\rm L}_{\rm Edd}$) and the spectral 
slope ($\Gamma$). We also let the normalization and slope ($A$ and $s$) of the power-law model for the positive (continuum) time-lags to be free parameters during the model fits but we kept 
them to be the same in all flux periods. In any case, the fits do not improve if we let either the normalization $A$ and/or the slope, $s$, to be different in each flux period.

\begin{figure*}
\centering
 \includegraphics[bb=41 -18 564 719,width=6.0cm,angle=270,clip]{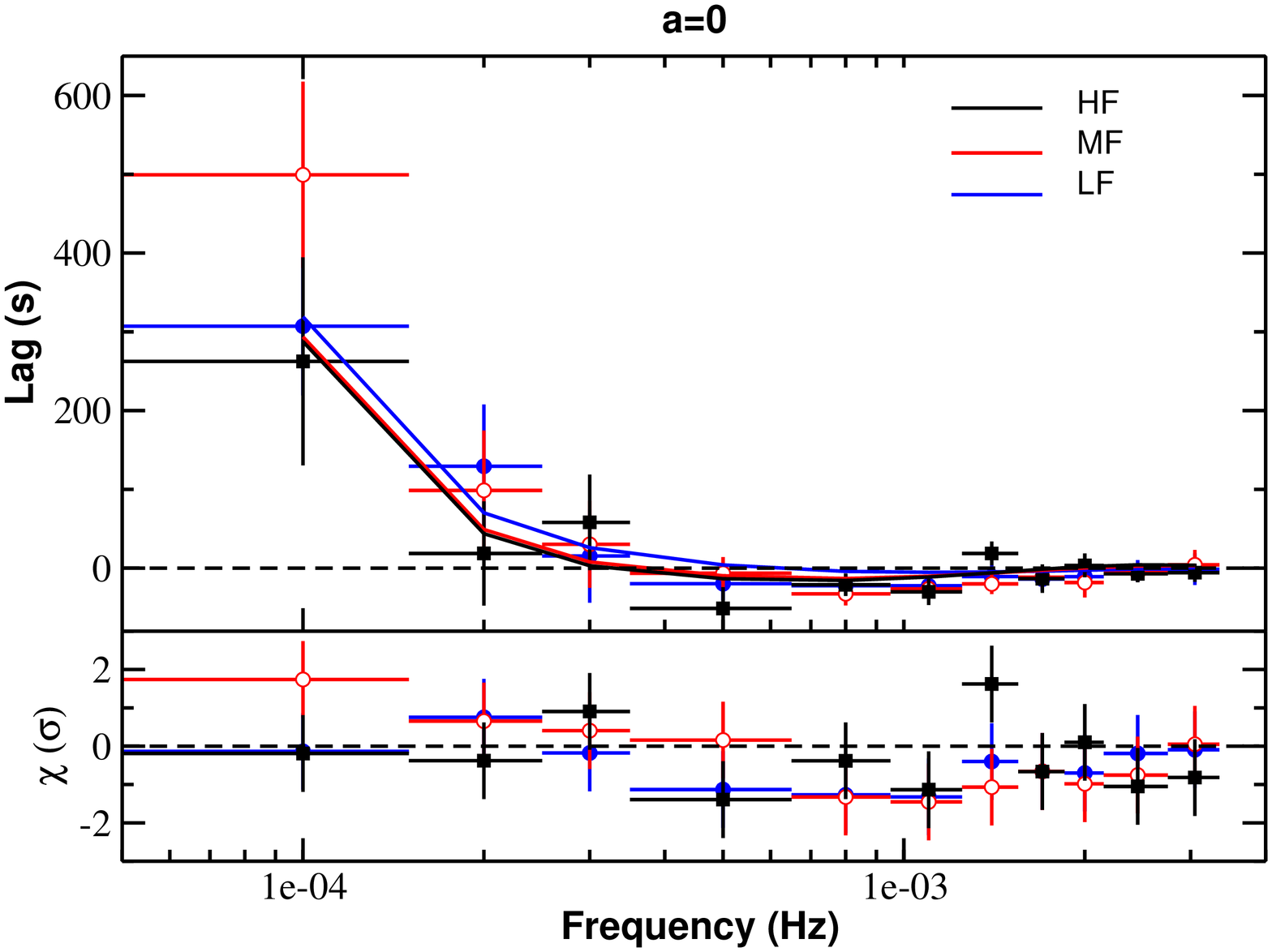}
 \includegraphics[bb=41 -18 564 719,width=6.0cm,angle=270,clip]{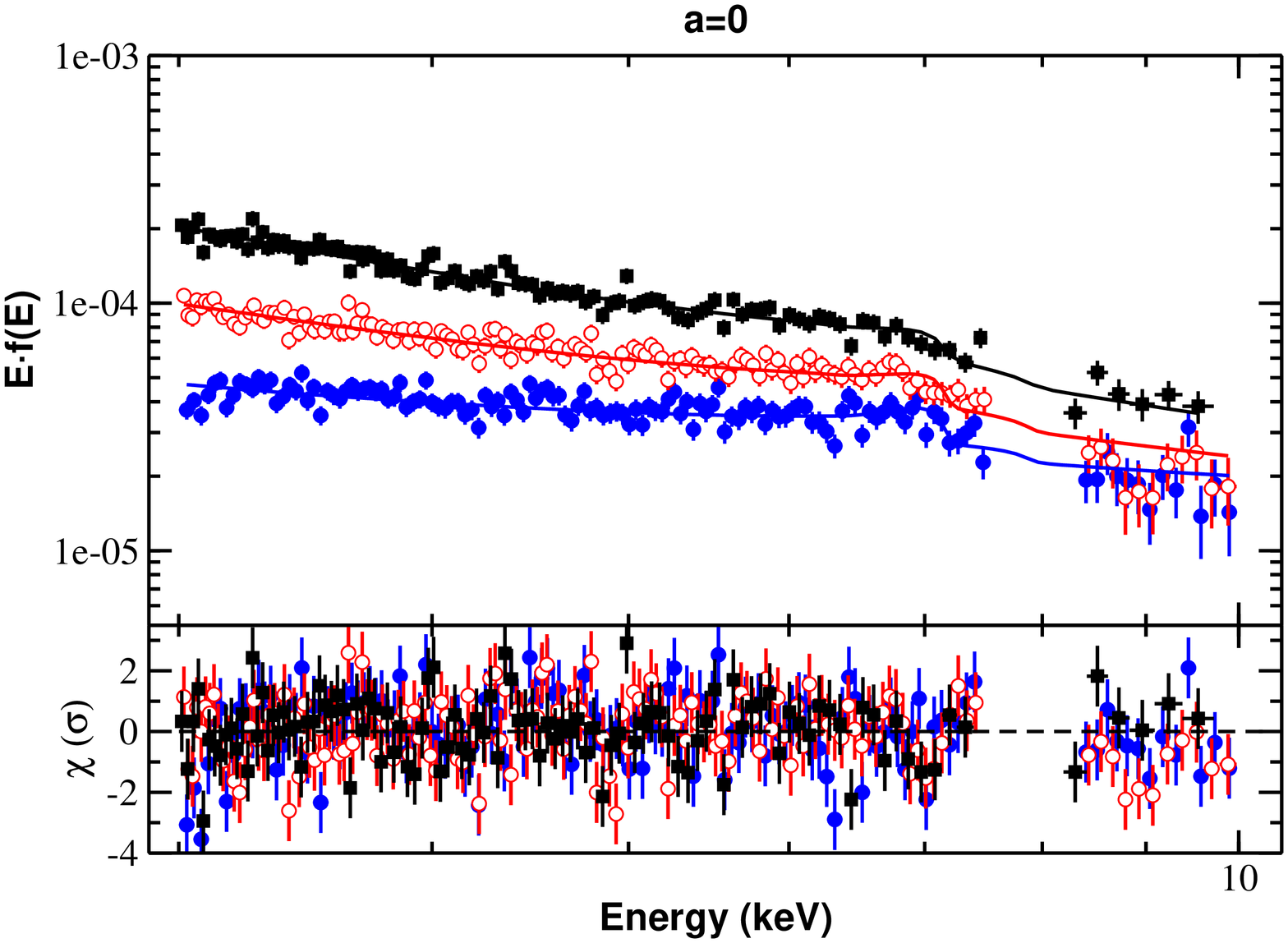}
 \includegraphics[bb=41 -18 564 719,width=6.0cm,angle=270,clip]{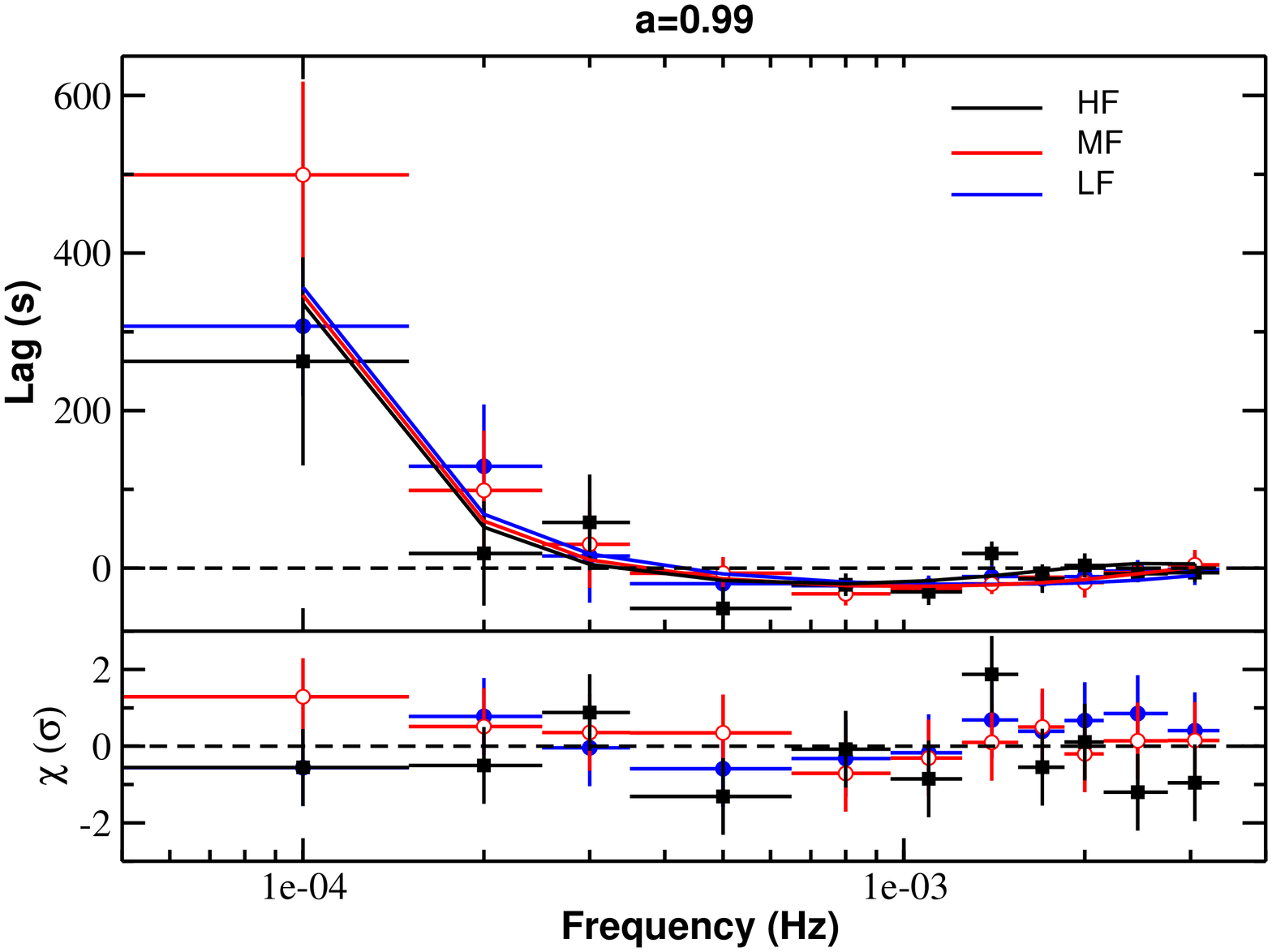}
 \includegraphics[bb=41 -18 564 719,width=6.0cm,angle=270,clip]{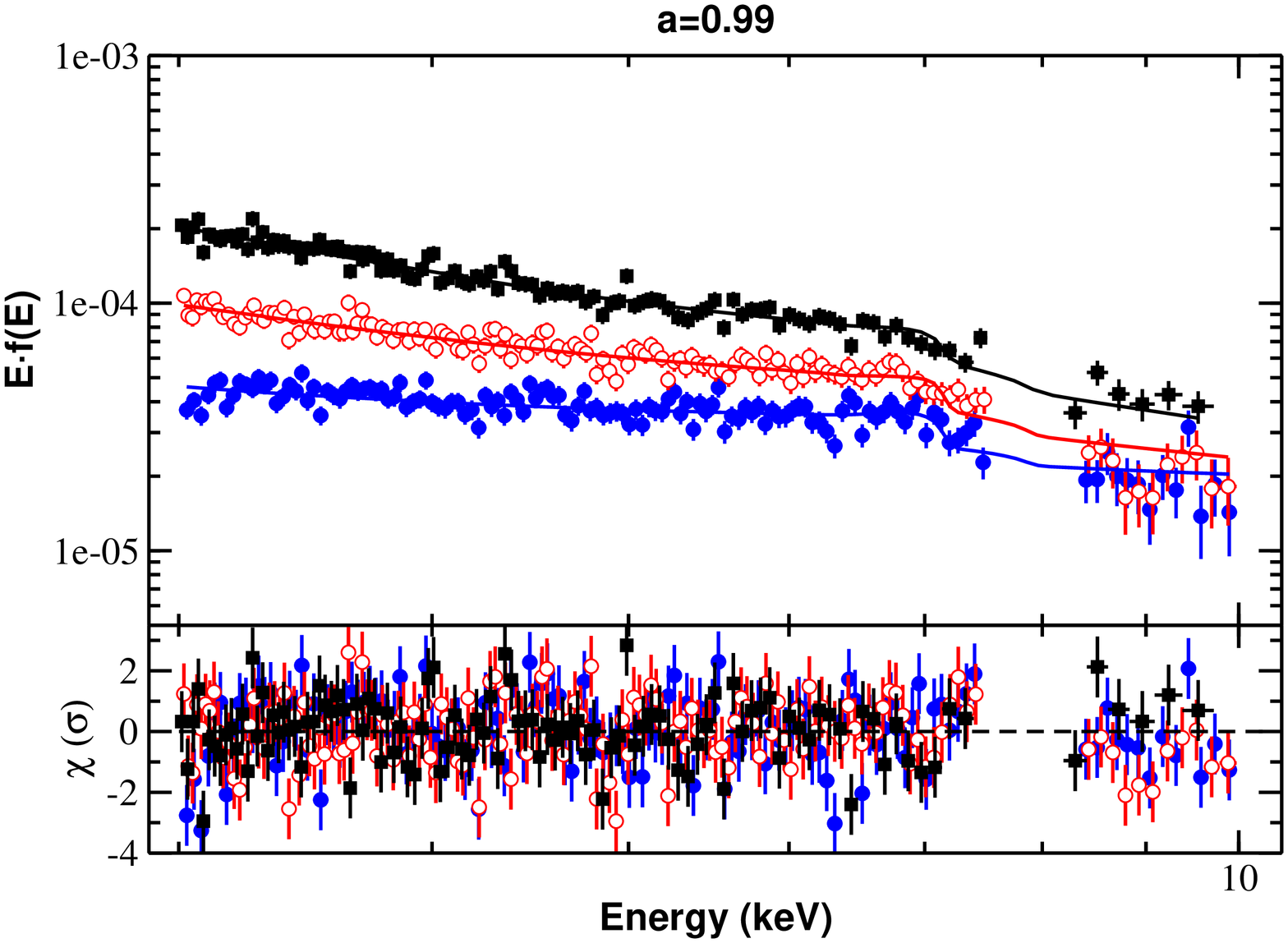}
\caption{The (0.3-1 versus 1-4\,{\rm keV}) X-ray soft time-lag versus frequency (left) and (3-10\,keV) energy spectra (right) fitted with the {\tt kynxilrev} and {\tt kynxillver} models, respectively, using the same values of the parameters obtained for spin zero (top) and rapidly spinning (bottom) from Tab.~\ref{log_results5}.  }
\label{fig:frequency}
\end{figure*}

\begin{table}
\begin{center}
  \caption{ Best-fit parameters using the models {\tt kynxilrev} and {\tt kynxillver} when we fit the time-lags and the energy spectra, respectively. We used a rapidly and zero spin values (left and right column, respectively). Errors indicate the 68\,\% confidence intervals.}
  \label{log_results5}
  \begin{tabular}{@{}lcc@{}}
  \hline
   $a$                                  &      $0.99$                           &    $0.0$        \\
\hline
\hline

 & Parameters common to  & \\
  & all three flux periods&  \\
  \hline
     ${\rm Z}/{\rm Z}_{\odot}$                        &    $4.9_{-0.4}^{+3}$                   &      $4.3_{-0.9}^{+0.8}$                                 \\
   Density\,($10^{15}\,{\rm cm}^{-3}$)              &    $2500_{-150}^{+1800}$               &      $2200_{-800}^{+900}$                                    \\
   ${\rm A (sec)}$                                &    ($4.5{\pm}1.0$)${\times}10^{-5}$    &      ($4.9{\pm}1.0$)${\times}10^{-6}$ \\
   ${\rm s}$                                        &    $1.7{\pm}0.3$\,(t)                       &      $1.9{\pm}0.4$\,(t)                                 \\
\hline 
               &           LF period                   &             \\
\hline
    $h$ (${\rm r_g}$)                 &    $3.4_{-0.7}^{+0.9}$                      &      $4.7_{-2.2}^{+2.3}$                  \\
   ${\Gamma}$                                       &    $1.91_{-0.06}^{+0.07}$                   &      $1.82_{-0.06}^{+0.04}$                            \\
   ${\rm L}_{2-10}/{\rm L}_{\rm Edd}$               &    ($1.44_{-0.10}^{+0.06}$)${\times}10^{-2}$  &      ($1.68{\pm}0.05$)${\times}10^{-2}$           \\
\hline
               &           MF period                                 &             \\
\hline
   $h$ (${\rm r_g}$)                &    $5.4_{-1.0}^{+1.7}$                      &      $8.4_{-2.9}^{+3}$                  \\
   ${\Gamma}$                                       &    $2.38_{-0.05}^{+0.04}$                   &      $2.32_{-0.06}^{+0.02}$                            \\
   ${\rm L}_{2-10}/{\rm L}_{\rm Edd}$               &   ($2.71_{-0.05}^{+0.08}$)${\times}10^{-2}$ &      ($2.92_{-0.03}^{+0.04}$)${\times}10^{-2}$           \\
\hline
               &           HF period                   &             \\
\hline
   $h$ (${\rm r_g}$)                &    $9.7_{-0.8}^{+10}$                       &      $9.7_{-8}^{+0.8}$                  \\
   ${\Gamma}$                                       &    $2.69_{-0.12}^{+0.04}$                   &      $2.65_{-0.05}^{+0.13}$                            \\
   ${\rm L}_{2-10}/{\rm L}_{\rm Edd}$               &    ($5.45{\pm}0.15$)${\times}10^{-2}$       &      ($5.41_{-0.21}^{+0.11}$)${\times}10^{-2}$           \\
\hline
   ${\chi}^{2}/{\nu}$                               &    $1.13$\,($469/414$)                      &   $1.18$\,($487/414$)                     \\
   p-value                                          &   $0.032$                                   &   $7.8e-3$                       \\
\hline
\hline
\end{tabular} 
\end{center}
 \footnotetext{ $^{1}$The normalization of the low frequency power-law time-lags, $A$, was kept tied in the fit. We used a value for the inclination of the accretion
disc of ${\theta}_{0}=45$\,deg.}
\end{table}

\subsection{The best-fit results}

We fitted, simultaneously, both the time-lags and the energy spectra of the three flux periods twice: once assuming $a=0$ and the second time fixing the spin parameter 
to 0.99. The model fitting of the time-lags and the energy spectra was done in {\tt XSPEC} 12.9.0 \citep{arnaud96}. The best fit results are reported in Tab.~\ref{log_results5} (middle and right 
columns for the $a=0.99$ and $a=0$ model fits, respectively). The best fit models are shown in Fig.~\ref{fig:frequency} (time-lags and energy spectra are plotted in the left and 
right panels, respectively). The cases of a Schwarzschild and a rapidly rotating BH are shown in the top and the bottom panels, respectively. 

Both the $a=0$ and the $a=0.99$ best-fit probability values are rather low. The poor quality of the model fits is mainly due to the best-fit residuals that appear 
at energies below $\sim 3.3$\,keV in the LF energy spectra (possibly due to the presence of a blue-shifted, S {\small XVI} absorption line at these 
energies, e.g. \citealt{jiang18}). Indeed, if we neglect the points below 3.3\,keV in the energy spectra the best-fit improves significantly, with 
$\chi^2=430.8$ and $439.8$ for 403 d.o.f. when $a=0.99$ and 0, respectively. We furthermore investigated if we can get a better fit by allowing the inclination to be a 
free parameter. The best fit inclination values are $47.5^{\rm o} \pm 1.8$ and $42^{\rm o}$$^{+2.5}_{-1.7}$ for the $a=0.99$ and 0 cases, respectively. These values 
are similar to the one we had originally assumed ($45^{\rm o}$). Since the fit does not improve significantly ($\Delta \chi^2 \sim 2-3$ for 414 d.o.f.) and the new model parameter values are 
consistent (within the errors) with the previous ones, we decided to keep the original best-fit 
results. 

Looking at the residuals of the best-fits to the time-lags (left column panels in Fig.~\ref{fig:frequency}), we can see 
that the $a=0$ model cannot fit the observed time-lags as good as the $a=0.99$ model. The best-fit $a=0$ model appears to be consistently overestimating the time-lags at 
frequencies higher than $\sim 3 \times 10^{-4}$\,Hz. As a result, $\Delta \chi^{2}=10$ between the best-fit of the $a=0.99$ and the $a=0$ model to the 
time-lags only ($\Delta \chi^{2}=18$ when we consider the energy spectra as well). It is difficult to quantify the improvement in the quality of the 
best-fit in the high-spin case, one reason being that the number of degrees of freedom (hereafter d.o.f.) is the same in both model fits. One way to do so is to compute  
the ``ratio of likelihoods", which is equal to $e^{\Delta \chi^2/2}$ (\citealt{mushotzky82}). In our case, this is identical to the ``evidence ratio", which is based on 
the Akaike \citep{akaike73} information criterion (see eq. 8 in \citealt{emmanoulopoulos16}). This ratio is a measure of the relative likelihood of one versus the 
other model, and is equal to $\sim 150$ in our case ($\sim 8100$ when we consider the energy spectra as well). This result does not mean that we can reject the 
hypothesis that the spin zero model is correct, but it does mean that the high spin model is much ``more likely" (i.e many 
hundreds to thousand times) to be the ``real/correct" model. We therefore conclude that our results strongly support the hypothesis of a rapidly-rotating BH
in {\iras}. 

The best-fit iron abundance and disc density values are consistent with the results from \citet{jiang18}, who performed a detailed spectral analysis of the 2016 observations using a similar
model to ours. The best-fit spectral slopes become steeper with increasing source flux, as commonly observed in AGN. Our results show that the height of the X-ray source 
in {\iras} is not constant. Instead, we find that the height should increase with increasing flux. According to 
our best-fit results, the X-ray source height increases from $\sim 4$ r$_g$ in the LF period to almost $\sim 10$ r$_g$ in the HF period. This trend of higher height 
of the corona with increasing X-ray luminosity is in agreement with the recent results of \citet{alston20}, to the time-lags only. The best-fit height values we find are systematically smaller 
than the best-fit results of \citet{alston20}. This is due to the fact that we use a different reflection model, and our best-fit disc density is almost 3 order of 
magnitude larger than the disc density that \citet{alston20} assumed in their model fitting. As a result, for the same X-ray luminosity, the disc in our case will be 
less ionized, and the reflection fraction will be smaller (for a given height). Consequently, the best-fit height is lowered, to achieve a reflection fraction similar to 
the one for a less dense disc and an X-ray source at a larger distance from the disc. 

\section{Discussion and conclusions} \label{sec:discussion}

We present the results from a combined study of the X-ray time-lag and energy spectra of {\iras}, which is one of the most variable and most observed NLSy1 AGN 
galaxies with {\it XMM-Newton}. For the timing analysis we used the data from EPIC-pn and the two MOS cameras on-board {\it XMM-Newton} to maximize the signal, whilst 
for the energy spectra we used the data from the EPIC-pn only. We used a new code for the computation and fitting of the time-lags spectra, {\tt KYNXILREV}. The code 
is publicly available, and it can be used to fit time-lags in {\tt XSPEC}. The main advantage over the previous code ({\tt KYNREFREV}, presented in G18) is that we 
use the recent {\tt XILLVER-D} tables \citep{garcia10,garcia13,garcia16}, and that we can let the disc density as a free parameter in the fit.

We estimated the time-lags between the 0.3-1 and 1-4\,keV bands, and we extracted the energy spectrum in the 3-10\,keV band. The time-lags are sensitive to 
the variability of the X-ray reflection component at low energies (i.e. of the variable soft excess), without being affected by the presence of an additional, black-body-like 
component, and of the warm absorber, as long as these components do not vary on short time scales. On the other hand, the energy spectrum in the 3-10\,keV band 
should be dominated by the main reflection feature in this band, i.e. the broad iron line. So, with our approach we can examine whether both the iron line and the 
soft-excess are consistent with the X-ray reverberation hypothesis, and constrain better the disc/corona properties.  

Moreover, we estimated and fitted (simultaneously) the time-lags and energy spectrum in three separate flux periods (low, medium and high-flux). To the best of our 
knowledge, this is the first time, for an AGN, that both the energy and the time-lags spectra are fitted in different flux periods. Our results are in agreement 
with the results of \citet{jiang18} and \citet{alston20}. We  find a positive correlation between the corona height and the continuum flux, like \citet{alston20}, and 
we show that the energy spectra in the three different flux states are fully consistent with the variation of the corona height with source flux. We find an electron 
disc density of $n_e \sim 2.5\times 10^{18}$ cm$^{-3}$ and an iron abundance of $\sim 5$ Z$_{\odot}$, like \citet{jiang18}, and we show that the time-lags are fully 
consistent with these disc properties. We find a best fit inclination of $\sim 45-50^{o}$, which is smaller than what \citet{jiang18} found, and more consistent with 
what is usually assumed for Type 1 AGNs. Finally, our model fitting results strongly suggest that the BH spin is close to 1 in \iras. 

The simultaneous fit of the timing/spectral properties constitutes a complete approach, and provides a ``unified" picture of the inner region in this AGN. Our work 
confirms that the results that have been reported from the study of either the energy or of the time-lags spectrum are indeed valid for both of them. Furthermore, by fitting 
simultaneously timing and spectral products, we can determine the disc/corona parameters more accurately. We believe that our results, together with the results from 
previous works (i.e. \citealt{chainakun16} and \citet{chainakun17}), reinforce the belief of the community that a combined timing/spectral fit can indeed provide a 
complete description of the processes, properties and the inner geometry of AGN. As new models and codes become publicly available, this possibility should start 
to be the common in the future. 

We discuss below some implications of the results we obtained from our analysis.

\subsection{The BH spin}

Following G18, we fitted the data by assuming either a rapidly spinning ($a=0.99$) or a non-rotating BH ($a=0$). The idea is to investigate whether the existing data can 
indicate that one option is significantly better than the other. G18 found that, by modelling the time-lags only of three AGN (excluding \iras\ ), it is not possible to 
discern between these two cases. A major result of our current study is that a (rapidly) spinning BH gives a significantly better fit to the time-lags and energy spectra. This is probably due to the 
fact that \iras\ has the best observations so far for this kind of X-ray reverberation studies. 

Although the model in the $a=0$ case can fit the energy spectra as well as the $a=0.99$ case (see the right panels in Fig.~\ref{fig:frequency}), it systematically underestimates the (negative) 
time-lags at frequencies above $\sim 3\times 10^{-4}$ Hz (see the upper left panel in the same figure). In fact the $a=0$ model cannot adjust freely the disc density and the corona height 
to fit the time-lags equally well as the $a=0.99$ solution, mainly due to the time-lag spectral feature. The energy
spectra, even though featureless, allowed us to find the values of the photon index and the observed luminosity for each flux period that helped us to well-constrain the final solution of the
fitting process. These results show the strength of the combined spectral/timing model fitting. The constraints of the model are very restrictive in our work, as we fit time-lags and energy 
spectra, simultaneously in three flux periods. Although we cannot estimate the BH spin with any accuracy, our results strongly indicate that the BH in {\iras} is rotating.

\subsection{The height of the X-ray source}

We find the corona height to increase with increasing X-ray luminosity. Our best fit results suggest an X-ray height of $\sim 3.5$ and $\sim 5.5$\,r$_{\rm g}$, when the observed 2-10\,keV 
X-ray luminosity is of $\sim 1.4$ and $\sim 2.7$ percent of the L$_{\rm Edd}$ (i.e. during the LF and MF periods, respectively). The height of the source increases to 
almost $\sim  10$\,r$_{\rm g}$ in the 
HF period, when the X-ray luminosity is $\sim 5.5$ percent of the Eddington limit. In fact, this change in the height of the source, if real, can explain (in part) the extreme observed X-ray 
variability amplitude of the source. We estimate the {\it intrinsic} X-ray luminosity to vary from $\sim 7$ to $\sim 9$ percent of the L$_{\rm Edd}$ (in the LF and HF period, respectively), which
is rather small compared to the observed light curve amplitude variation.  

IRAS 13224-3809 is the first AGN where flux-related time-lags variations were detected \citep{kara13}. In this work we confirm the previous claim by \citet{kara13} that the X-ray source may be 
located closer to the accretion disc in the lower-flux than during the high flux periods. Our best model-fits suggest that flux related height variations can explain the time-lags 
in various flux periods, similar to what \citet{kara13} proposed for the time-lags variations during flaring and quiescent periods. Our results are fully consistent with \citet{alston20}, who 
studied the same source, using (almost) the same observations. These authors studied the time-lags only, using a previous version of the X-ray reverberation code that the one we use in 
this work. They estimated the time-lags using the data from single {\it XMM-Newton} orbits, and they found that the time-lags are best fitted when the height of the X-ray corona increases 
with increasing X-ray flux. We studied the time-lags simultaneously with the energy spectra, in three broader flux periods. The loss in time ``resolution" of the time-lags evolution in 
our study is somehow counter-balanced by the use of the energy spectra. In both studies the results are the same: assuming the lamp-post hypothesis, the height of the X-ray source increases 
by a factor of $\sim 3$ when the observed X-ray luminosity appears to increase by factor of $\sim 4$.

The height of the X-ray corona appears to be $\sim 3-5$ r$_{\rm g}$ when the source is in the LF and MF period, but then it increases by a factor of two (at least) when the source is in 
the HF period. Fig.~\ref{fig:lc1-4} shows that the source spends less time in the HF period, which is characterized by very fast, ``spiky", flare-like events. These are similar to 
the ``spiky" flares in the simulations of \citet{ghisellini04}, in the case of the ``aborted jet" scenario for the X-ray emission in radio-quiet AGN (see for example the figures 4 and 7 in 
their paper). According to these authors the X-ray emission is due to ``blobs" of material, which can reach a maximum radial distance and then fall back, colliding with the blobs produced 
later and still moving outwards. Possibly the large source height during the HF periods may be due to collisions that happened at higher heights either because the blobs have a larger initial 
velocity or they are launched at larger distances from the BH (both assumptions result in the same effect, see \S 5.2 in \citealt{ghisellini04}). At the same time, if the width of these 
blobs increases as they move (due to expansion) their dissipation time scale will also increase (eq. 23 in \citealt{ghisellini04}), which could also explain the decrease on the characteristic 
break frequency in the PSD of the source with increasing source flux \citep{alston19}.

We furthermore note that the best-fit residuals of the HF time-lags have a wavy-like appearance. The wavy-residuals are not significant (the best-fit $\chi^2$ to 
the time-lags of the spectrum in the HF period is statistically acceptable), but we find their presence intriguing. G18 had also observed wavy-residuals in the time-lags spectra of the three 
AGN they had studied, as well as \citet{chainakun17} in the case of PG 1244+026. These residuals cannot be explained in the case of the simple lamp-post geometry, but 
as \citet{chainakun17} suggested, they might be indicative of the presence of two, co-axial point sources. This is an interesting possibility, which could suggest 
that, when in the HF state, both a large and lower height sources may co-exist in \iras, with the lower height source being fainter. An alternative explanation for these 
wavy residuals was recently proposed by \citet{chainakun19}. These authors considered that the time-lags would be produced in the case of a spherical corona, and they showed
that the bumps and  wiggles in the time-lags should be present in this case. They also found stronger wiggles when the corona is optically thicker, or when it becomes 
larger. It might be that the corona expands in \iras\ as its height increases, producing this wavy pattern in the observed time-lags spectra.

\section*{Acknowledgments}

MCG, MD and VK acknowledge the support by the project RVO:67985815 and GACR project 18-00533S. MCG acknowledges funding from ESA through a partnership with IAA-CSIC (Spain).

\section*{Data availability}

\noindent The {\tt KYNXILREV} and {\tt KYNXILLVER} codes used to fit the data are publicly available at:\\ \indent https://projects.asu.cas.cz/stronggravity/kynreverb/ 
\\ and \\ \indent https://projects.asu.cas.cz/stronggravity/kyn/ \\ All the data presented in this paper are available upon (reasonable) request.

\bsp




\end{document}